\documentclass[%
 reprint,
 amsmath,amssymb,
 aps,
]{revtex4-2}

\usepackage{graphicx}% Include figure files
\usepackage{dcolumn}% Align table columns on decimal point
\usepackage{bm}% bold math
\usepackage{braket}
\usepackage{xcolor}
\begin{document}

\preprint{APS/123-QED}

\title{Comparative analysis of diverse methodologies for portfolio optimization leveraging quantum annealing techniques}% Force line breaks with \\

\author{Zhijie Tang}
\email{zhijietang0710@163.com}%Lines break automatically or can be forced with \\

\author{Alex Lu Dou}%
 % \email{alex.dou@pwc.com}
 \affiliation{%
Innovation Hub, PricewaterhouseCoopers AC Shanghai}%
\author{Arit Kumar Bishwas}%
\email{(Corresponding author)aritkumar.official@gmail.com}
\affiliation{%
Innovation Hub, PricewaterhouseCoopers USA}%

\begin{abstract}
Portfolio optimization (PO) is extensively employed in financial services to assist in achieving investment objectives. By providing an optimal asset allocation, PO effectively balances the risk and returns associated with investments. However, it is important to note that as the number of involved assets and constraints increases, the portfolio optimization problem can become increasingly difficult to solve, falling into the category of NP-hard problems. In such scenarios, classical algorithms, such as the Monte Carlo method, exhibit limitations in addressing this challenge when the number of stocks in the portfolio grows. Quantum annealing algorithm holds promise for solving complex portfolio optimization problems in the NISQ era. Many studies have demonstrated the advantages of various quantum annealing algorithm variations over the standard quantum annealing approach. In this work, we conduct a numerical investigation of randomly generated unconstrained single-period discrete mean-variance portfolio optimization instances. We explore the application of a variety of unconventional quantum annealing algorithms, employing both forward annealing and reverse annealing schedules. By comparing the time-to-solution(TTS) and success probabilities of diverse approaches, we show that certain methods exhibit advantages in enhancing the success probability when utilizing conventional forward annealing schedules. Furthermore, we find that the implementation of reverse annealing schedules can significantly improve the performance of select unconventional quantum annealing algorithms.
\end{abstract}

\maketitle

%\tableofcontents

\section{\label{sec:level1}Introduction}

The portfolio optimization problem is a well-known finance challenge that aims to maximize the possible returns while minimizing risk. The objective of solving this problem is to determine an optimal asset allocation that  meets the specified requirements. However, the complexity of the portfolio can escalate to NP-hard levels as the number of involved assets and constraints increases~\cite{erwin2023meta,suthiwong2016cardinality}. Classical algorithms, including the commonly used Monte-Carlo algorithm, may fail to provide efficient solutions due to the computational complexity associated with the portfolio optimization problem. 

Quantum computing has benefits in solving certain hard problems where classical algorithms fail~\cite{shor1994algorithms,grover1996fast,lloyd1996universal,abrams1997simulation,shor1999polynomial,farhi2000quantum,steffen2003experimental,aspuru2005simulated,montanaro2016quantum,farhi2016quantum,von2021quantum,mohseni2017commercialize,cao2019quantum,orus2019quantum,bishwas2021managing,bishwas2022parts}. Its applicability extends to many disciplines, including artificial intelligence~\cite{schuld2015introduction,bishwas2016big,bishwas2018all,bishwas2020investigation,bishwas2020gaussian,caro2022generalization,cerezo2022challenges,zeguendry2023quantum}, finance~\cite{hegade2022portfolio,albareti2022structured,herman2023quantum}, the optimization of complex systems~\cite{ajagekar2019quantum,ajagekar2020quantum,abel2022quantum}, and drug discovery~\cite{cao2018potential,zinner2021quantum,cova2022artificial,mahesh2023accelerating}. The fundamental operating units in quantum computers are essentially different from their classical counterparts. Quantum computing leverages the phenomenon of quantum entanglement, which enables extensive parallel processing by allowing the simultaneous operation of multiple states. Additionally, quantum superposition provides a vast storage space compared to classical bits. These unique characteristics of quantum computing offer the potential for exponential advancements in computational power and storage capabilities, surpassing the limitations of classical computing systems. The gate-model quantum computing, which employs quantum logical gates for information processing, is the most extensively studied quantum computing model and has been proven to be universal computation~\cite{deutsch1985quantum}. However, the fragility of qubits poses a significant challenge in scaling up quantum gate model computers in practice applications. Quantum error correction and fault tolerance techniques that are required in the implementation of quantum gate model computer can lead to large resource overhead costs~\cite{shor1995scheme,calderbank1996good,steane1996error,fowler2012surface,devitt2013quantum,lidar2013quantum,roffe2019quantum}. As a result, current quantum gate model computers are limited to a few hundred qubits~\cite{gambetta2020ibm}. As an alternative to the quantum gate model, quantum annealing is extensively used in solving optimization problems.  The scale of quantum annealing hardware can reach thousands of qubits~\cite{king2023quantum,mcgeoch2023milestones}. The high scalability of quantum annealer enables more practical implementations of quantum computation. 

The most commonly used techniques in portfolio optimization problems, such as mean-variance model~\cite{naik2023portfolio}, Value-at-Risk~\cite{jorion1997value,jorion2007value}, and Conditional Value-at-Risk~\cite{rockafellar2000optimization}, have been extensively studied with classical methodologies, including statistical and artificial intelligence approaches~\cite{milhomem2020analysis,zhang2020deep,sen2021stock,yashaswi2021deep,gunjan2023brief,al2023implementing}. The investigations on portfolio optimization utilizing quantum computing have been conducted with gate-based models, employing techniques such as the quantum approximate optimization algorithm (QAOA)~\cite{hodson2019portfolio,barkoutsos2020improving,herman2022portfolio,vesely2022application}, the variational quantum eigensolver (VQE)~\cite{barkoutsos2020improving,liu2022layer,mugel2022dynamic}, the quantum algorithm for linear systems of equations(HHL)~\cite{yalovetzky2021nisq,vesely2022application}, and the quantum annealing algorithm~\cite{rosenberg2015solving,marzec2016portfolio,elsokkary2017financial,venturelli2019reverse,orus2019quantum,cohen2020picking,cohen2020portfolio,cohen2020portfolio2,grant2021benchmarking,hegade2022portfolio,mugel2022dynamic,lang2022strategic,palmer2022financial,mattesi2023financial}. Portfolio optimization problems involving diverse financial assets can lead to an exponential number of variables. Gate-based models are constrained in addressing the challenges due to the hardware limitations. Consequently,  quantum algorithms employed for portfolio optimization problems predominantly rely on quantum annealing in the NISQ era, even when considering the limited connectivity of current quantum annealing hardware. 

In the context of quantum annealing, a portfolio optimization problem can be encoded as a problem Hamiltonian. The system evolves from the ground state of an easily prepared initial Hamiltonian to the ground state of the problem Hamiltonian. According to the adiabatic theorem, if the system's evolution is sufficiently slow, the final ground state attained by the system will encode the solution to the optimization problem. Various adaptations of QA have demonstrated advantages over the conventional QA~\cite{seki2012quantum,seki2015quantum,hormozi2017nonstoquastic,susa2017relation,susa2018quantum,susa2018exponential,kapit2021noise,tang2021unconventional,chen2010fast,del2013shortcuts,guery2019shortcuts}. The standard QA involves a stoquastic Hamiltonian, wherein the off-diagonal elements are real and non-positive. A stoquastic Hamiltonian that avoids “sign problem” can be efficiently simulated using classical algorithms like Quantum Monte Carlo(QMC). Thus, it’s valuable to explore some variants of standard QA that employ non-stoquastic Hamiltonians. Additionally, the reverse annealing protocol has been proven to provide quantum advantages~\cite{ohkuwa2018reverse,venturelli2019reverse,passarelli2020reverse}. Therefore, it is also worthwhile to explore variants of QA that incorporate the reverse annealing schedule.

Recent research lacks a comprehensive study that compares different variants of QA in the domain of solving portfolio optimization problems. We believe that such a study has the potential to provide valuable guidance for the experimental implementation of QA. To address this research gap, we propose conducting an in-depth investigation into the portfolio optimization problem utilizing a carefully selected ensemble of unconventional QA algorithms. Additionally, we intend to conduct a comparative analysis of the performance of each algorithm, utilizing both forward annealing and reverse annealing. 

The paper is structured as follow: in Section \ref{Portfolio optimization formulation} we introduce the formulation of the portfolio optimization problem and the recent developments. Section \ref{Unconventional QA methods} describes different promising variants of QA like adding transverse couplers to standard QA, inhomogeneous driving QA, RFQA and CDQA. Section \ref{Reverse annealing} introduces the reverse annealing schedule. And section \ref{Results and Discussion} is devoted to the numerical results, we compare the time-to-solution and success probability of each algorithm, utilizing both forward annealing and reverse annealing. In Section \ref{Conclusion} we conclude with a discussion and considerations for future work.

\section{Portfolio optimization formulation}\label{Portfolio optimization formulation}
Portfolio optimization is a challenging problem in finance and management. Its objective is to find a balance between risk and return. For instance, the general formulation of Markowitz's mean-variance model can be expressed as follows~\cite{barkoutsos2020improving}:
\begin{equation}
 \max_{x} \left\{ \sum_{i}^{n} \mu_i x_i - q \sum_{i,j}^{n} \sigma_{ij} x_i x_j - \lambda \left( B-\sum_{i}^{n} x_i\right)^2 \right\}, 
 \label{eq1}
\end{equation}
% Eq.~(\ref{eq1})
the given formulation is a constrained mean-variance model, where $\mu_i$ is the expected return of assets, $\sigma_{ij}$ is the risk of assets, and $B$ is the full budget. 

The formulation of a particular portfolio optimization depends on the given market conditions. To adapt a portfolio optimization problem for quantum annealing, it is necessary to transform the formulation into a Quadratic Unconstrained Binary Optimization (QUBO) problem, where the variables are limited to values of 0 or 1. The transformation can be made through various conversions, such as binary, unary, or sequential encoding.~\cite{tamura2021performance,rosenberg2015solving}. The problem is encoded into a problem Hamiltonian $H_P$, which is defined by pauli matrices
\begin{equation}
H_P\ =\ \sum_{i}^{n}h_i\sigma_i^z\ +\ \ \sum_{i,j}^{n}J_{ij}\sigma_i^z\sigma_j^z\ +\ C ,
\label{eq2}
\end{equation}
where $h_i$ is the qubit biases, $J_{ij}$ is the coupling strength between qubit $i$ and $j$, $C$ is a constant generated from the transformation and can be omitted in the following expressions. Unlike the transverse field ising model(TFIM)~\cite{kadowaki1998quantum}, which considers the nearest neighbour interactions, the Ising spin-glass model derived from the portfolio optimization problem encompasses all-to-all connections. Current quantum annealer such as the D-Wave machine has limited connectivity due to the QPU topology. However, as the hardware technology advances, we believe it is worthwhile simulating the behavior of promising quantum annealing algorithms for future implementation.

Quantum-based algorithms have demonstrated their efficacy in enhancing portfolio optimization outcomes, such as the works using gate-model Quantum computing in~\cite{barkoutsos2020improving,yalovetzky2021nisq,herman2022portfolio,plekhanov2022variational,lim2022quantum,liu2022layer,vesely2022application,wang2022classically,brandhofer2022benchmarking} and the ones using Quantum annealing algorithms in~\cite{rosenberg2015solving,elsokkary2017financial,venturelli2019reverse,cohen2020picking,cohen2020portfolio,cohen2020portfolio,phillipson2021portfolio,hegade2022portfolio,lang2022strategic,palmer2022financial,mattesi2023financial,xu2023dynamic}. Despite the limited connectivity of quantum annealers, they provide more accessible stable qubits than gate-model quantum computer in the NISQ era. Consequently, our focus in this study will be on evaluating the performance of different quantum annealing algorithms. To make a comprehensive comparison of QA algorithms for solving the portfolio optimization problem, we transform the portfolio optimization model into an ising model and simulate the annealing process under the guidance of time dependent Schrodinger equation. Various adaptations of QA algorithms are studied and compared to provide guidance for their implementation on real quantum hardware. Furthermore, the reverse annealing of each algorithm is  investigated and introduced in the following sections.

\section{Unconventional QA methods}\label{Unconventional QA methods}
%  to ``float'',
The formulation of standard quantum annealing is as follows,
\begin{equation}
H(t)\ =\ A(t)H_0+B(t)H_P,
 \label{eq3}
\end{equation}
the initial Hamiltonian $H_0$ and problem Hamiltonian $H_P$ are not commute with each other, the annealing parameters $A(t)$ and $B(t)$ follows the routine that $A(t)\gg B(t)$ at the beginning and $B(t)\gg A(t)$ at the end. In the forward annealing routine, $A(t)$ monotonically decreases while $B(t)$ monotonically increases. The initial Hamiltonian, conventionally chosen as $H_0=-\sum_{i=1}^{n}h\sigma_i^x$, has an easily prepared ground state. Its ground state $\ket{+}^{\bigotimes n}$ is the uniform superposition state, where $\ket{+}=\ \frac{1}{\sqrt2}(\ket{0}+\ket{1})$. 
The problem Hamiltonian is usually prepared with pauli-z matrices so that $H_P$ doesn't commute with $H_0$. The ground state of $H_P$ encodes the solution to the computational optimization problem. 

In standard quantum annealing, the total Hamiltonian interpolates between the initial Hamiltonian and problem Hamiltonian under slowly adiabatic evolution, the system will stay on the instantaneous ground state according to the adiabatic theorem. The total annealing time is proportion to the inverse of squared minimum energy gap $t_f\propto \Delta_{min}^2$ , the probability of staying on the ground state can be approximated as $P_{t_f}=\ 1-e^{-\alpha t_f}$~\cite{zener1932non}. For problems that go through first order transition, the evolution time required to reach a reasonable success probability will grow exponentially with the system size. However, it's not ideal to solving realistic problems with an exponentially long evolution time. Some techniques that can provide speedup on short time evolution emerges for practical use requirements. We introduce them in the following subsections.

\subsection{Coupler QA}
In the work of~\cite{seki2012quantum,seki2015quantum,hormozi2017nonstoquastic,susa2017relation}, adding a 2-body transverse coupler is proven to boost the traditional quantum annealing algorithm in solving some classes of problem. The formulation of the total Hamiltonian is given as:
\begin{equation}
H(t)\ =(1-s(t))H_0+s(t)H_P+(1-s(t))s(t)H_I.
 \label{eq4}
\end{equation}
The coupler term can be ferromagnetic coupler $H_I^F$, antiferromagnetic coupler $H_I^A$, or a mixed coupler $H_I^M$. They are defined as follows

\begin{equation}
  \begin{aligned}
H_I^F=-\sum_{<i,j>}^{N}{\sigma_i^x\sigma_j^x} \\
H_I^A=+\sum_{<i,j>}^{N}{\sigma_i^x\sigma_j^x} \\
H_I^M=\sum_{<i,j>}^{N}{r_{ij}\sigma_i^x\sigma_j^x},
  \end{aligned}
 \label{eq5}
\end{equation}
where $r_{ij}$ can be a random number chosen from $\{-1,1\}$. The coupler term disappears at the beginning and the end of the annealing process, ferromagnetic coupler keeps the stoquastic of the tranditional QA, anti-ferromagnetic and mixed coupler forms non-stoquastic Hamiltonians. The work in~\cite{hormozi2017nonstoquastic} shows that adding transverse coupler outperforms the tranditional QA. Studies on constructing non-stoquastic Hamiltonians with $H_I^F$ or $H_I^M$ have shown the enhanced capabilities of the transverse couplers~\cite{seoane2012many,seki2012quantum,seki2015quantum}, they can help to avoid the first order phase transition in some certain systems.

\subsection{Inhomogeneous driving QA}
In addition to adding transverse coupler to the transverse driving field, an inhomogeneous driving field can weaken the effect of phase transition and lead to a better performance~\cite{susa2018quantum,susa2018exponential} compared with traditional QA, the formulation is given by
\begin{equation}
H(t)\ =H_V(t)+s(t)H_P,
 \label{eq6}
\end{equation}
unlike the uniform transverse field in standard QA, the inhomogeneous driving Hamiltonian $H_V(t)$ can be defined as 

\begin{equation}
H_V(t) = -\sum_{i=1}^{N(1-\tau)} \sigma_i^x, 
 \label{eq6_1}
\end{equation}
where $\tau$ is another annealing controller similar to $s(t)$ that starts from 0 and ends up with 1 so that the transverse field on each qubit will be turned off starting from site $N$ to site 1. The $H_V(t)$ term can also be represented by a continuous function $\Gamma_i(t)$ as follow
\begin{equation}
H_V(t) = -\sum_{i=1}^{N}\Gamma_i(s)\sigma_i^x,
 \label{eq6_2}
\end{equation}
the continuous filed amplitude $\Gamma_i(t)$ is applied to each qubit and will be reduced sequentially. The definition of $\Gamma_i(t)$ given in ~\cite{susa2018exponential} is
\begin{equation}
  \begin{split}
  &\Gamma_i(s) =
    \begin{cases}
      1 & \text{if $s<s_i$}\\
      N(1-s^r) + (1-i) & \text{if $s_i \leq s \leq s_{i-1}$}\\
      0 & \text{if $s_{i-1}<s$}\\
    \end{cases} 
  \end{split}
 \label{eq6_3}
\end{equation}
where $s_i = (1-\frac{i}{N})^{\frac{1}{r}}$ and $r=0.5$ is chosen in our work. Such a continuous function can be reduced to Eq.\ref{eq6_1} as $N \to \infty$ and will circumvent the sudden change in Hamiltonian.

Experimental implementation and analytical studies of inhomogeneous driving QA have shown its ability to enhance the success probability for some specific instances~\cite{hartmann2019quantum,adame2020inhomogeneous}

\begin{figure*}
\centering
  \includegraphics[width=10.8cm,height=7.2cm]{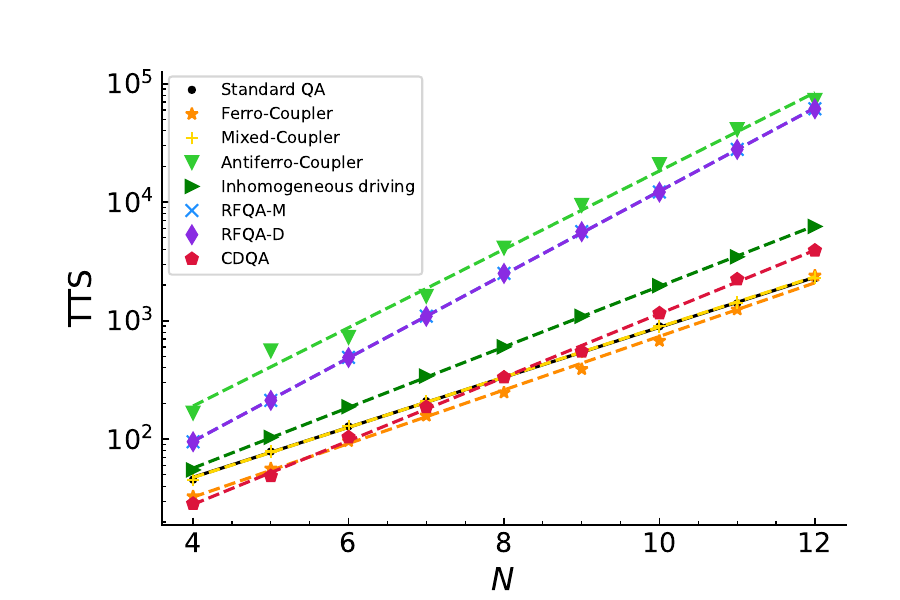}
\caption{ Shown is the time required to identify the true ground state in randomly generated instances using various unconventional QA methods with a forward annealing schedule. The runtime is $t_f=1$. The data points corresponding to the standard QA method are depicted as black dots, while the black solid line represents the best-fit curve for the data. Other markers and dashed lines are used to represent other methods for comparison purpose. The presented results are the average of 500 randomly generated instances.}
\label{fig_tts}
\end{figure*}

\begin{figure*}
\centering
  \includegraphics[width=20cm,height=12cm]{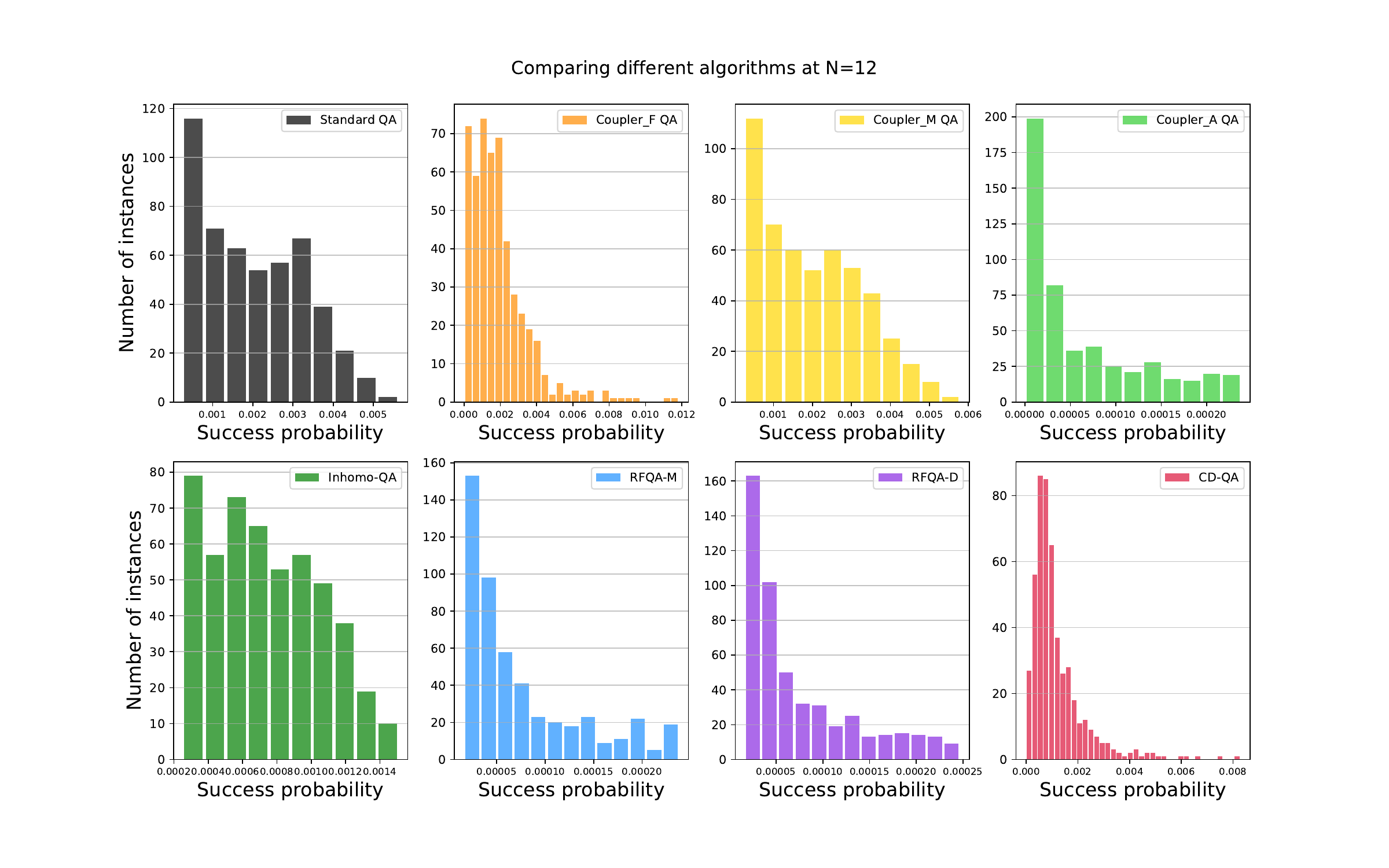}
\caption{A comparison of the ground state success probability of different algorithms employing the forward annealing schedule at N=12. For each algorithm, 500 random instances were generated.}
\label{fig_bin}
\end{figure*}

\subsection{RF-QA}
Traditional QA finds the ground state by adiabatically evolve the system to avoid the diabatic transitions nearby the energy gap. RFQA method is a new variant of QA that enhance the probability of finding ground state by taking advantage of diabatic transitions. RFQA modifies the conventional uniform transverse field into oscillating field and leave the problem Hamiltonian unchanged, the modified field can be either transverse field with oscillating magnitude or a uniform field with oscillating direction in x-y plane. RFQA method can be implemented in current hardware by making some modifications to the control circuitry~\cite{kapit2021noise}.

The RFQA field with oscillating magnitude is refered to as $RFQA-M$, $M$ stands for magnitude. The formulation of the driving field in RFQA-M is denoted by 
\begin{equation}
H_M=-\sum_{i}^{N}{(1+\bar{\alpha_i}sin(2\pi f_it))\sigma_i^x},
 \label{eq7_m}
\end{equation}
in which $\bar{\alpha_i}$ represents the amplitude of the field, and the frequency $f_i$ is randomly chosen and applied to each spin $i$, the low frequency oscillating field induces multi-photon resonances during the annealing process, the proliferating multi-photon resonances can rapidly mix the ground state and first excited states at the adjacent of the minimum gap, and provides a polynomial speedup over the traditional quantum annealing algorithm theoretically~\cite{kapit2021noise}.

The oscillating field can also be realized by oscillating the direction in x-y plan, which we refer to $RFQA-D$ method, $D$ represents direction here, the field in RFQA-D is denoted by 
\begin{equation}
H_D=-\sum_{i}^{N}[cos(\bar{\alpha_i}isin(2\pi f_i t))\sigma_i^x + sin(\bar{\alpha_i}isin(2\pi f_i t))\sigma_i^y].
 \label{eq7_d}
\end{equation}
The Hamiltonian in RFQA method is defined as follow
\begin{equation}
H(t)\ =(1-s(t))H_{M/D}+s(t)H_P.
 \label{eq7}
\end{equation}

The advantages of RFQA have been shown in~\cite{tang2021unconventional}, which shows a polynomial quantum speedup of RFQA in solving an artificial problem with competing ground states. 

\begin{table}[b]
\centering
\begin{tabular}{p{5cm}p{3cm}}
 \hline
 \hline
Algorithm & Scaling exponent\\
\hline
\\
Standard QA & 0.70  \\
\\
Ferro Coupler QA & 0.75  \\
\\
Mixed Coupler QA & 0.70 \\
\\
Anti-ferro Coupler QA & 1.10 \\
\\
Inhomogeneous QA & 0.85  \\
\\
$RFQA_M$ & 1.17  \\
\\
$RFQA_D$ & 1.16  \\
\\
CD-QA & 0.89  \\
 \hline
\end{tabular}
\caption{The scaling exponents for each method employ the forward annealing schedule. The $TTS(N)$ of each approach is modeled by fitting it into the $2^{\beta+\alpha N}$ function. We present the exponential scaling coefficient ``$\alpha$" for each method in the table for comparison purposes.}
\end{table}

\subsection{CD-QA}
In standard QA, the ground state of $H_P$ can be found with high probability after annealing time that relates to the minimum energy gap. In some problems with exponentially small energy gap, the required annealing time can be extremely long.  To make quantum annealing more practical to use, the idea of suppressing nonadiabatic transition during short time evolution was proposed~\cite{chen2010fast,del2013shortcuts,guery2019shortcuts}, the non-adiabatic technique is named as counter-diabatic quantum annealing (CD-QA). The effectiveness of CD-QA has been proven in the works~\cite{bason2012high,del2012assisted,zhang2013experimental,ban2014counter,okuyama2016classical,claeys2019floquet,hartmann2020multi,prielinger2021two,funo2021general,hegade2021shortcuts,wurtz2022counterdiabaticity,chandarana2022digitized,hegade2022portfolio,hegade2022digitized,keever2023towards,chandarana2023digitized,chandarana2023meta,vcepaite2023counterdiabatic}

The idea of CD-QA is taking advantage of the Adiabatic Gauge Potential(AGP) to suppress the non-adiabatic transition in quantum annealing. By transforming the original Hamiltonian $H_{origin}(t)$ in a specific rotating frame that consists of a static effective Hamiltonian and an adiabatic gauge potential related Hamiltonian which contribute to the excitations between eigenstates, we can add a Hamiltonian to $H_{origin}(t)$ so that the quantum state
keeps adiabaticity and evolves to the ground state in a shorter time.

We can formulate the total Hamiltonian in CD-QA as follow(Appendix \ref{appendix1}):
\begin{equation}
H(t)\ =H_{origin}(t)\ +\ H_{CD},
 \label{eq16}
\end{equation}
the counter-diabatic driving term can be chosen as:
\begin{equation}
H_{CD}=\dot{\lambda}\sum_{i}^{N}{\alpha_i\left(t\right)\sigma_i^y}.
 \label{eq16_1}
\end{equation}
The annealing schedule $\lambda(t)$ can be chosen as $\lambda(t) ={sin}^2[{\frac{\pi}{2}sin}^2(\frac{\pi t}{2T})]$ so that the CD driving term vanishes at $t=0$  and $t=t_f$.
And $\alpha_i(t)$ specifically for the ising model is derived in Appendix \ref{appendix1}.

Ideally, one can evolve the system in a short time without excitations between the eigenstates. This concept is widely used in many fields such as quantum physics, stochastic physics. 

\section{Reverse annealing}\label{Reverse annealing}
\begin{table}[b]
\centering
\begin{tabular}{p{5cm}p{3cm}}
 \hline
 \hline
Algorithm & Scaling exponent\\
\hline
\\
Standard QA & 1.00  \\
\\
Reverse Standard QA & 0.34  \\
\\
Reverse Ferro Coupler QA & 0.35  \\
\\
Reverse Mixed Coupler QA & 0.34  \\
\\
Reverse Anti-ferro Coupler QA & 0.35 \\
\\
Reverse $RFQA_M$ & 0.09  \\
\\
Reverse $RFQA_D$ & 0.09  \\
 \hline
\end{tabular}
\caption{The scaling exponents for certain methods employ the reverse annealing schedule and the standard QA employ the forward annealing schedule. The $TTS(N)$ of each approach is modeled by fitting it into the $2^{\beta+\alpha N}$ function. We present the exponential scaling coefficient ``$\alpha$" for each method in the table for comparison purposes.}
\end{table}

In conventional forward quantum annealing, the transverse field is initially activated and subsequently reduced as the annealing process progresses, allowing the system to explore the entirety of the solution space. For most practical NP-hard problems that are described by the Ising model, the conventional routine can be doomed by exponentially small gaps in the spin-glass phase~\cite{knysh2016zero}. Reverse annealing that takes advantage of partial knowledge on the solution space can offer the potential for more efficient exploration of the true solution. The idea of reverse annealing is starting from a candidate initial state that is obtained by other methods, such as classical algorithms. The transverse field strength is initialized at $s=1$ and subsequently decreased to a finite value, followed by a pause. Finally, it is ramped up again to $s=1$. The system starts the exploration of the solution space from a candidate state nearby the solution in such a reverse annealing routine, thereby enhancing the efficiency of the search process. The advantages of reverse annealing are revealed in some works~\cite{chancellor2017modernizing,ohkuwa2018reverse,perdomo2019readiness}.
Reverse annealing can be described by the follow formulation
\begin{equation}
H(t)\ =\ sH_p+(1-s)H_0,
 \label{eq}
\end{equation}
where $H_p$ and $H_0$ represent the problem and initial Hamiltonians as in conventional forward annealing QA. However, the annealing parameter $s$ differs from that used in conventional forward annealing QA. 

\begin{figure}[ht]
\centering
\includegraphics[width=8.55cm,height=6.3cm]{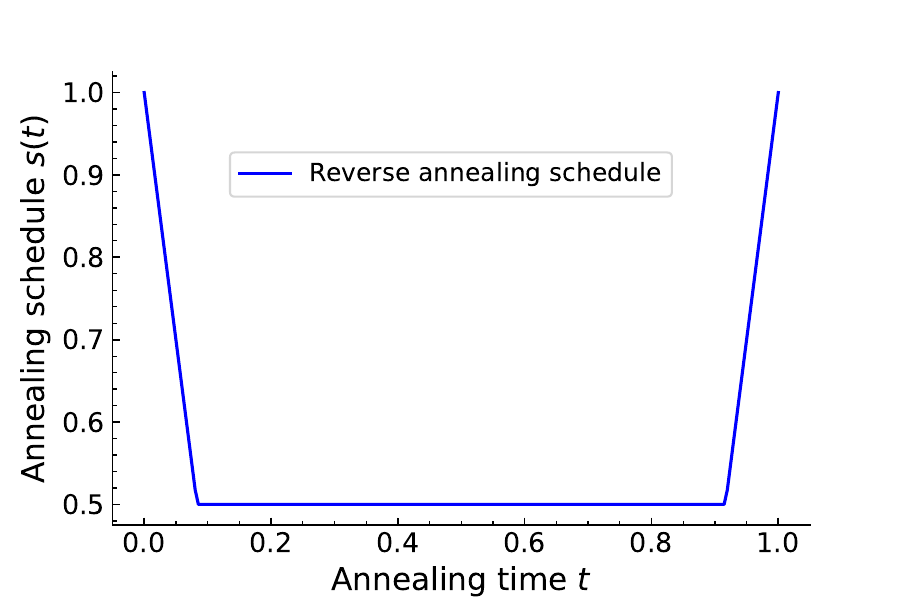}
\caption{The reverse annealing schedule $s(t)$ initiates at 1, where the problem Hamiltonian dominates. Following this, $s(t)$ is gradually reduced to 0.5 and remains at this value for a relatively long time before being increased back to 1.}
\label{reverse_schedule}
\end{figure}

\begin{figure}[ht]
\centering
\includegraphics[width=8.55cm,height=6.3cm]{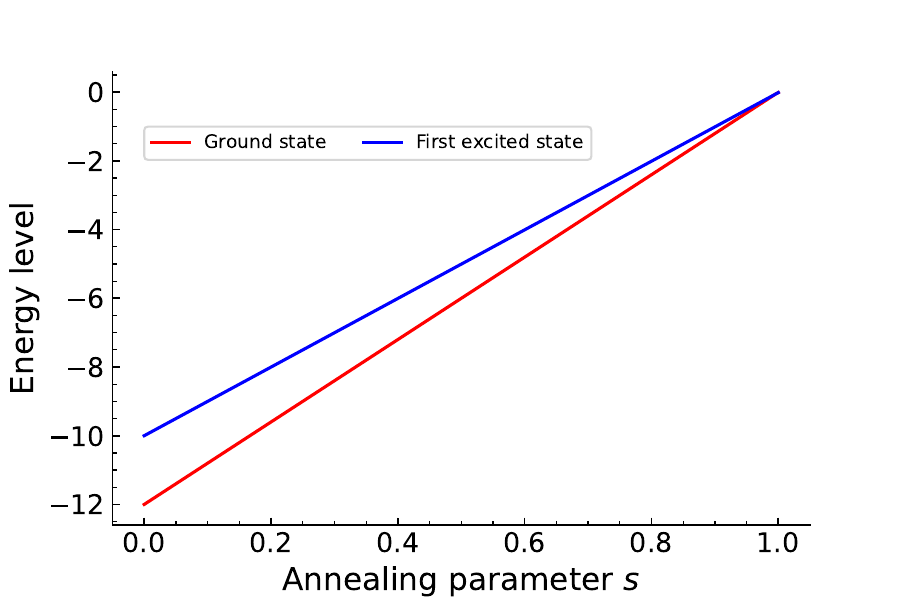}
\caption{The energy levels of the hard instance at $N=12$. The ground state is represented by the red solid line, while the first excited state is represented by the blue line. An avoided small energy gap appears near the end of the annealing schedule.}
\label{energy}
\end{figure}

\begin{figure}[ht]
\centering
  \includegraphics[width=8.55cm,height=6.3cm]{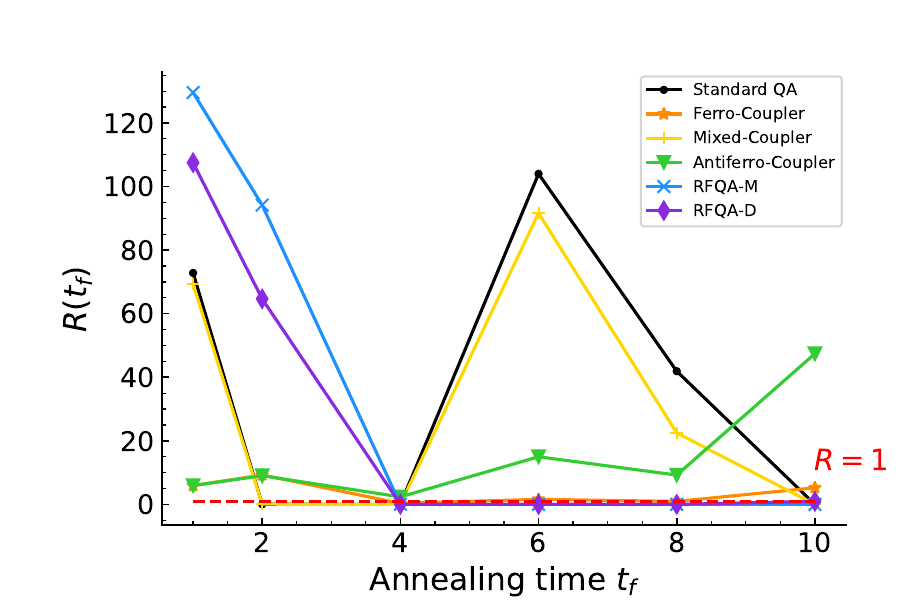}
\caption{The advantages ration $R(t_f)$ of the hard instance at $N=12$. The red dashed line represents $R=1$, indicating whether an algorithm achieves performance advantages with the help of reverse annealing schedule. The other solid lines correspond to the advantages ratio of each algorithm.}
\label{ratio}
\end{figure}

The reverse annealing schedule in this work is illustrated in FIG.~\ref{reverse_schedule}, the annealing parameter $s$ starts at 1, ramps down to $s=0.5$ and pause at $s=0.5$ for a long time, and then ramp up to $s=1$. 

\begin{figure*}
\centering
  \includegraphics[width=10.8cm,height=7.2cm]{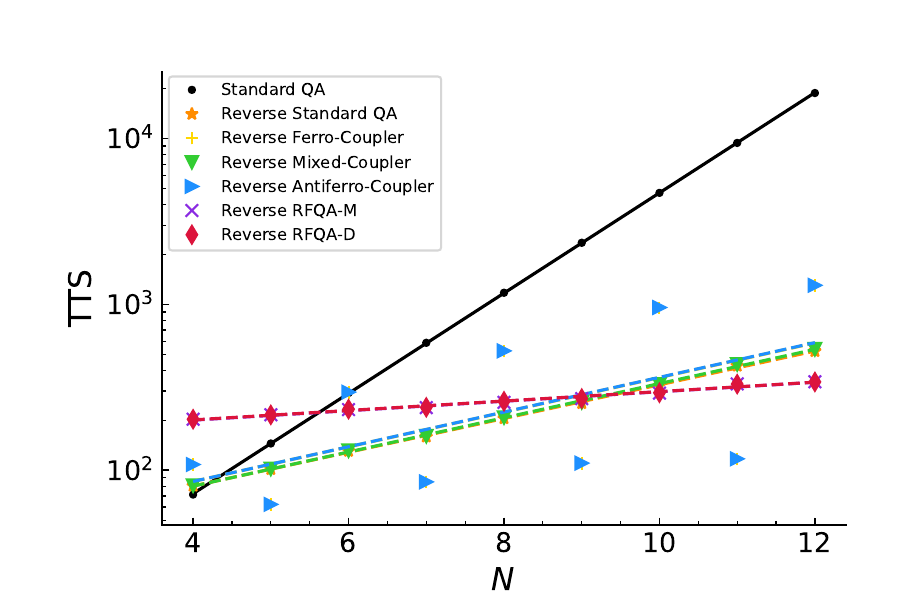}
    \caption{Shown is the time required to find the true ground state in randomly generated instances using various unconventional QA methods with a reverse annealing schedule. The runtime is $t_f=1$. The data of the standard QA method are denoted by black dots, and the black solid line represents the best-fit curve of the data. Other markers and dashed lines are used to represent the results of other methods for comparison purpose. The results presented are averaged over 500 randomly generated instances.}
\label{re_fig_tts}
\end{figure*}

\begin{figure*}
\centering
\includegraphics[width=20cm,height=12cm]{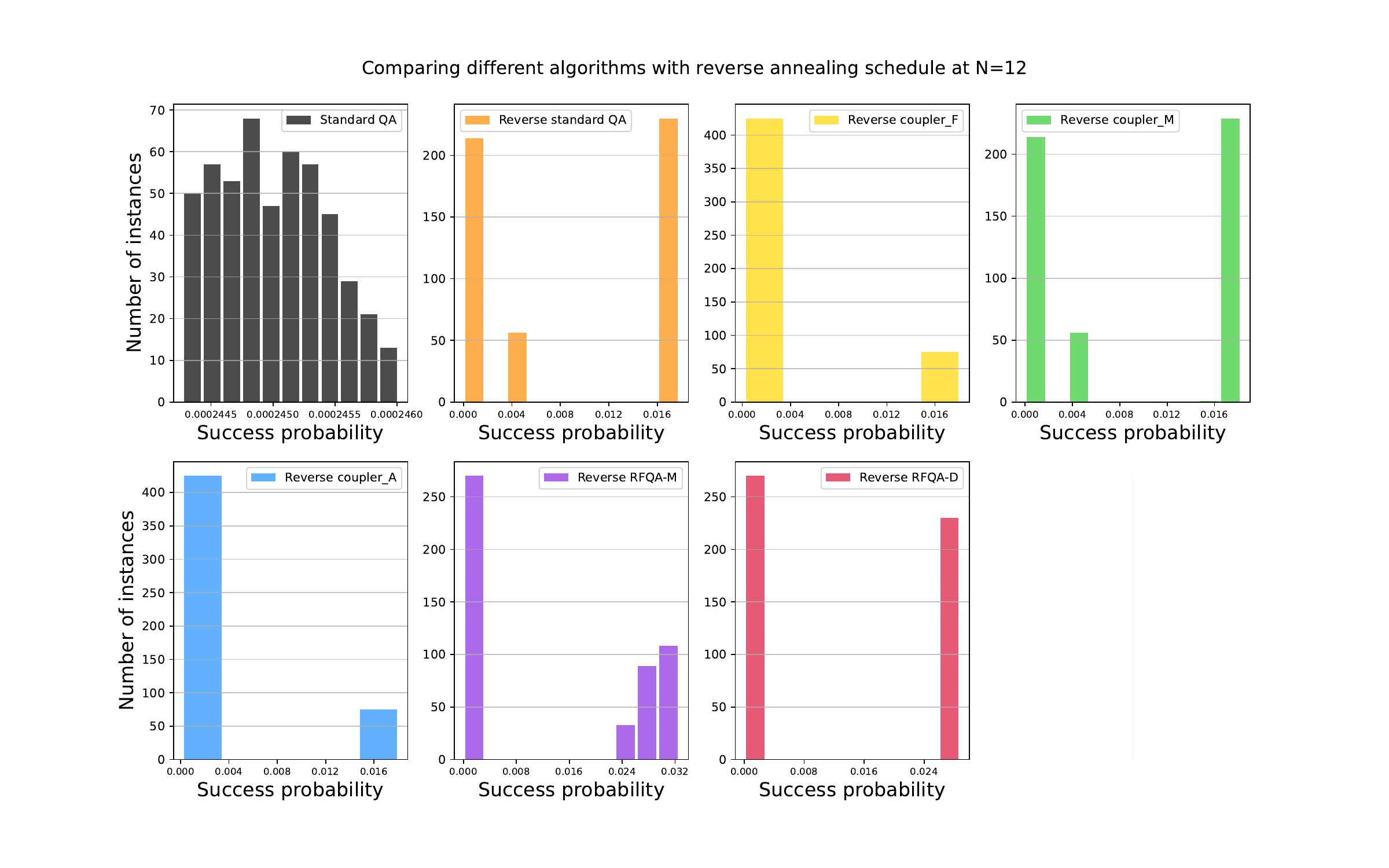}
\caption{A comparison of the ground state success probability between certain methods that use the reverse annealing schedule and the standard QA that employs the forward annealing schedule at N=12. For each algorithm, 500 random instances were generated. }
\label{re_fig_bin}
\end{figure*}

\section{Results and Discussion}\label{Results and Discussion}
% Although quantum computer hardware is rapidly developed, the price of accessing a quantum hardware is still  expensive. Considering the high cost of running a quantum algorithm on a hardware. We simulate the process of quantum annealing with a classical computer, the quantitative results generated by the simulation is meaningful to provide a realistic impact in solving optimization problems. 
In this section, we present and discuss the quantitative results obtained from a performance comparison of alternative QA approaches, employing both forward and reverse annealing schedules. These findings aims to provide guidance on the experimental implementation of these approaches. 
% The total annealing time is $t_f=1$ and time step is set as $dt=0.05$
 
In the study employing the forward annealing schedule, we generate 500 random instances for the unconstrained single-period discrete portfolio optimization problem. For each method, we determine the success probability within an annealing time $t_f=1$. The system size ranges from $N=4$ to $N=12$. In addition to the success probability, we compute another performance metric called time-to-solution(TTS) to evaluate the algorithms' performance. TTS calculates the time needed to obtain a solution with 99\% success probability: 
\begin{equation}\label{TTS}
TTS \propto t_f \frac{\ln(1-0.99)}{\ln(1-p(t_f))}.
\end{equation}

As illustrated in FIG.~\ref{fig_tts}, the randomly generated instances for portfolio optimization demonstrate exponential complexity, as the time-to-solution grows exponentially with system size $N$. We see that the unconventional QA algorithms do not exhibit obvious advantages over standard QA approach. To make a more intuitive comparison, we fit the $TTS(N)$ of each method with an exponential function $2^{\beta+\alpha N}$ and list the extracted exponent $\alpha$ in TABLE I. It is evident that  only the mixed coupler QA approach is comparable to the standard QA. However, this doesn't imply that the unconventional QA algorithms lack value in addressing portfolio optimization problems. In FIG.~\ref{fig_bin}, we present the success probability of the 500 instances and demonstrate that methods such as the Ferromagnetic coupler QA, Mixed coupler QA and CDQA can enhance the success probability in certain cases. 

% Reverse annealing requires a prior knowledge to the solution space. 

In the study employing the reverse annealing schedule, we find one hard instance that contains a local minimum and we designate this local minimum as the initial state for the reverse annealing schedule. The energy gap of the hard instance is depicted in FIG.~\ref{energy}, revealing the presence of an avoided energy gap between the first excited state and the ground state towards the end of the annealing schedule. This avoided gap is extremely small, thereby inducing diabatic transitions between the two competing states. To show how the advantages provided by reverse annealing changes with annealing time, we conducted a comparative analysis of the  success probability for some algorithms annealed using both forward and reverse annealing schedules. This analysis was performed over a range of short annealing times, specifically from $t_f=1$ to $t_f=10$, as depicted in FIG.~\ref{ratio}. We define the success probability of each algorithm as $P_f(t_f)$ for the forward annealing schedule and $P_r(t_f)$ for the reverse annealing schedules. To quantify the advantages of the reverse annealing schedule in each algorithm, we denote the advantage as the ratio
\begin{equation}
R(t_f) = \frac{P_r(t_f)}{P_f(t_f)}.
 \label{eq_ratio}
\end{equation}
The advantages ratio for each algorithm is denoted as $R_S(t_f)$, $R_F(t_f)$, $R_M(t_f)$, $R_A(t_f)$, $R_{RM}(t_f)$, $R_{RD}(t_f)$, where ``S'', ``F'', ``M'', ``A'', ``RM'', and ``RD'' represent standard QA, Ferro Coupler QA, Mixed Coupler QA, Anti-ferro Coupler QA, RFQA-M, and RFQA-D, respectively. For example, $R_S(t_f)$ is the ratio of the standard QA using the reverse annealing schedule to the standard QA using the forward annealing schedule. As shown in FIG.~\ref{ratio}, the dashed red line at $R=1$ represents the scenario where the reverse annealing schedule does not exhibit any advantages. It's evident that all the methods demonstrate significant advantages in most cases. Reverse annealing that starts from a known possible solution provides obvious help in enhancing the success probability of finding the true ground state. Furthermore, we studied how the quantum advantages provided by reverse annealing changes with system size from $N=4$ to $N=12$. The time-to-solution (TTS) comparison between algorithms employing the reverse annealing schedule and the standard QA approach using the forward annealing schedule is depicted in FIG.~\ref{re_fig_tts}. Additionally, the TTS scaling exponent for each algorithm is listed in TABLE II. The findings presented in FIG.~\ref{re_fig_tts} and TABLE II indicate that all methods, with the assistance of reverse annealing, exhibit promising quantum advantages in solving challenging portfolio optimization problem sets when compared to the standard QA approach utilizing the forward annealing schedule. Specifically for the system size at $N=12$, we conduct a comparison of the success probability between the standard QA approach using the forward annealing and other methods employing the reverse annealing schedule, as depicted in FIG.~\ref{re_fig_bin}. 500 random instances are generated for the purpose of comparison. Notably, the methods utilizing the reverse annealing schedule exhibit significant improvements in the success probability for certain instances. 

\section{Conclusion}\label{Conclusion}
In this work, we investigated the portfolio optimization problem using different promising unconventional QA algorithms. We show that certain algorithms have the capability to improve the success probability compared to the standard quantum annealing algorithm in specific cases. Furthermore, we observed that the advantages can be further augmented with the help of the reverse annealing schedule. 

In the context of the forward annealing schedule, we generated problem instances randomly and compared the time-to-solution exponent and success probability of each method. The simulation results reveal that while the unconventional QA methods may not exhibit obvious advantages in improving the averaged time-to-solution, certain methods such as coupler-QA and CD-QA demonstrate an enhanced success probability compared to the standard QA approach.

In the investigation with reverse annealing, we examined the behavior of standard QA, Coupler-QA and RFQA with the reverse annealing schedule. Specifically, we explored how the advantages ratio of reverse annealing changes with annealing time on a specific hard instances characterized by an extremely small avoided energy gap. FIG.~\ref{ratio} indicates that reverse annealing effectively enhances the success probability of standard QA, Coupler-QA and RFQA in most cases, particularly at short annealing times. The improvement in success probability is further illustrated in FIG.~\ref{re_fig_bin}. Moreover, we compared the time-to-solution scaling of each methods using reverse annealing. The findings presented in FIG.~\ref{re_fig_tts} and TABLE II clearly indicate the significant quantum advantages provided by reverse annealing. 

In conclusion, the utilization of unconventional QA methods has demonstrated the potential to enhance the success probability in certain cases. Moreover, the application of the reverse annealing schedule has significantly improved the performance of selected methods. We believe that the comprehensive quantitative assessment of time-to-solution and success probability for each method serves as a crucial step towards their experimental implementation in the NISQ era. It is worth noting that the portfolio optimization model employed in this study is the Markowitz portfolio model, where risk is estimated using the standard deviation. For future work, exploring alternative metrics such as the Sharpe ratio and CVaR could be considered.

% \begin{acknowledgments}
% ZJ.T thanks Arit kumar bishwas and Alex Lu Dou for useful discussions. We acknowledge the support of PWC \\
% \end{acknowledgments}

\appendix
\section{CD-QA Hamiltonian Derivation}
\label{appendix1}
To suppress the nonadiabatic transition, the adiabatic gauge potential (AGP) $A_{\lambda}$ is introduced into the Hamiltonian H in the lab frame and transform the Hamiltonian into a stationary effective Hamiltonian ${H^\prime}_{eff}$ in a rotating frame~\cite{kolodrubetz2017geometry,bhattacharjee2023lanczos}.  For the state $\ket{\psi}$ in the lab frame, the corresponding state $\ket{\widetilde{\psi}}$ in the rotating frame becomes $\ket{\widetilde{\psi}} =\ U^{\dag}(\lambda(t))\ket{\psi}$, $U$ is the unitary transformer between the frames and $\lambda$ is the parameter that controls the evolution of $H$. The evolution of the state in rotating frame is guided by the time dependent Schrodinger equation
\begin{equation}
i\hbar\frac{d\ket{\widetilde{\psi}}}{dt}=H_{eff}\ \ket{\widetilde{\psi}}= (U^{\dag}HU -\ i\hbar\dot{\lambda}U^{\dag}\partial_{\lambda}U )\ket{\widetilde{\psi}}.
 \label{eq8}
\end{equation}
The first term in the right-hand side is a diagonal term and only the second term contributes to the excitations between eigenstates. By adding the adiabatic gauge potential $A_\lambda= i\hbar\partial_{\lambda} U U^{\dag}$ , the corresponding effective Hamiltonian becomes $U^{\dag} HU$ and the state in the rotating frame becomes stationary state, the nonadiabatic transition is largely suppressed as a result. 

The limitation of this approach is that the exact AGP is always consists of nonlocal terms, which is challenging to implement with the current quantum hardware. Thus, approximate methods for finding optimal CD driving protocols that can be easier to implement with local operations are proposed in~\cite{sels2017minimizing}. In the ising model $H_{Ising}\ =\ \sum_{i=1}^{n}h_i\sigma_i^z\ +\ \ \sum_{<i,j>}^{n}J_{ij}\sigma_i^z\sigma_j^z$, the simplest local counter-diabatic term can be represented by the magnetic field along y direction: $H_{CD}=\dot{\lambda} A_\lambda^\ast$ . To find the approximate AGP term $A_\lambda^\ast=\sum_{i}^{N}{\alpha_i\left(t\right)\sigma_i^y}$, an equivalent way of Eq.~(\ref{eq9}) is minimizing the Hilbert-Schmidt norm of $G_\lambda$
\begin{equation}
G_\lambda(A_\lambda^\ast)\ =\ \partial_\lambda H\ +\ \frac{i}{\hbar}[A_\lambda^\ast,H].
 \label{eq10}
\end{equation}
And this can be converted to minimize the action $S(A_\lambda^\ast)\ =\ Tr[{G_\lambda}^2(A_\lambda^\ast)]$ of $A_\lambda^\ast$~\cite{sels2017minimizing}
\begin{equation}
\frac{\delta S(A_\lambda^\ast)}{\delta A_\lambda^\ast}=0.
 \label{eq11}
\end{equation}

The adiabatic gauge potential satisfies
\begin{equation}
[i\hbar\partial_\lambda H\ -\ [A_\lambda,H],H] = 0,
 \label{eq9}
\end{equation}
\\
the total Hamiltonian H(t) is given by
\begin{equation}
\begin{aligned} 
H(t) &= (1-\lambda(t))\sum_{i=1}^{n}h_x\sigma_i^x \\
& +\lambda(t)(\sum_{i=1}^{n}h_i^z\sigma_i^z+ \sum_{i,j=1}^{n}J_{ij}\sigma_i^z\sigma_j^z),
 \label{eq12}
\end{aligned}
\end{equation}
where $h_x=-1$ by convention. We denote the coefficient of $\sigma_i^x$,$ \sigma_i^z$ and coupling term $\sigma_i^z\sigma_j^z$ as $X_i$,$Z_i$,$C$,where $X_i=(1-\lambda(t))h_x, Z_i =\lambda(t)h_i^z ,C=\lambda(t)J_{ij} $. By substituting the expression of $A_\lambda^\ast$ and the total Hamiltonian $H(t)$ into Eq.~(\ref{eq12}), the Hermitian operator now is

\begin{equation}
\begin{aligned} 
G_\lambda(A_\lambda^\ast) &=\sum_i \sum_{n\neq i} [(\dot{X}_i-2\alpha_i Z_i)\sigma_i^x + ( \dot{Z}_i+2\alpha_i X_i)\sigma_i^z \\
& + \dot{C} \sigma_i^z \sigma_n^z -2 \alpha_i C (\sigma_i^x \sigma_n^z+\sigma_i^z \sigma_n^x)].
 \label{eq13}
 \end{aligned}
\end{equation}
The dot of coefficient represents the derivative with respect to $\lambda$. The Hilbert-Schmidt norm of $G_\lambda(A_\lambda^\ast)$ is simply adding up the squares of all coefficients

\begin{equation}
\begin{aligned}
2^{-N}Tr[{G_\lambda}^2(A_\lambda^\ast)] &= \sum_i \sum_{n\neq i} [(\dot{X}_i-2\alpha_i Z_i)^2 \\
& +(\dot{Z}_i+2\alpha_i X_i)^2+\dot{C}^2+8\alpha_i^2 C^2 ].
 \label{eq14}
 \end{aligned}
\end{equation}

We obtain the optimal AGP by minimizing the above equation, which gives the expression of $\alpha_i$~\cite{sels2017minimizing,hartmann2019rapid}

\begin{equation}
\begin{aligned}
\alpha_i(t) &=\frac{\dot{X}_i Z_i-\dot{Z}_i X_i}{2({{\ Z}_i}^2+{{\ X}_i}^2+2C^2)} \\
&= -\frac{h_x h_i^z \dot{\lambda}}{2[h_x^2(\lambda(t)-1)^2+{\lambda(t)}^2 (h_i^{z^2}+2\sum_{i\neq j}J_{ij}^2)]} .
 \label{eq15}
 \end{aligned}
\end{equation}
Then we derive the formulation of the CD-QA in Ising model as follow:
\begin{equation}
H(t)\ =(1-\lambda(t))H_0+\lambda(t)H_{Ising}\ +\ \dot{\lambda}\sum_{i}^{N}{\alpha_i\left(t\right)\sigma_i^y}.
 \label{eq_final}
\end{equation}
\nocite{*}

\bibliography{bib}% Produces the bibliography via BibTeX.

\end{document}